\documentclass[apjl]{emulateapj}
\usepackage{apjfonts}
\usepackage{graphicx}

\newcommand{\cmjj}{\mbox{${\rm cm^{-2}}$}}
\newcommand{\hI}{\mbox{${\rm H\,I}$}}

\newcommand{\ha}{\mbox{${\rm H}\alpha$}}
\newcommand{\apg}{\gtrsim}
\newcommand{\apll}{\lesssim}
\newcommand{\etal}{\ensuremath{\mbox{et~al.}}}

\newcommand{\ibid}{\underline{\makebox[0.5in]{}}.}

\newcommand{\ewr}{\mbox{$W_r(2796)$}}

\shorttitle{Mg\,II Absorbing Gas around Galaxies}
\shortauthors{Chen \etal}

\begin{document}

\slugcomment{Accepted for Publication in the Astrophysical Journal Letters}

\title{What Determines the Incidence and Extent of Mg\,II Absorbing Gas Around Galaxies?}

\author{Hsiao-Wen Chen\altaffilmark{1}, Vivienne Wild\altaffilmark{2}, Jeremy L.\ Tinker\altaffilmark{3}, Jean-Ren\'e Gauthier\altaffilmark{1}, Jennifer E.\ Helsby\altaffilmark{1}, Stephen A.\ Shectman\altaffilmark{4}, \& Ian B.\ Thompson\altaffilmark{4}}

\altaffiltext{1}{Dept.\ of Astronomy \& Astrophysics and Kavli Institute for Cosmological Physics, University of Chicago, Chicago, IL, 60637, U.S.A.
{\tt hchen@oddjob.uchicago.edu}}
\altaffiltext{2}{Institut d'Astrophysique de Paris, CNRS, Universit\'e Pierre \& Marie Curie, UMR 7095, 98bis bd Arago, 75014 Paris, France}
\altaffiltext{3}{Berkeley Center for Cosmological Physics, University of California-Berkeley}
\altaffiltext{4}{The Observatories of Carnegie Institution for Science, 813 Santa Barbara St., Pasadena, CA 91101}

\begin{abstract}

  We study the connections between on-going star formation, galaxy
  mass, and extended halo gas, in order to distinguish between
  starburst-driven outflows and infalling clouds that produce the
  majority of observed Mg\,II absorbers at large galactic radii ($\apg
  10\ h^{-1}$ kpc) and to gain insights into halo gas contents around
  galaxies.  We present new measurements of total stellar mass
  ($M_{\rm star}$), \ha\ emission line strength (${\rm EW}(\ha)$), and
  specific star formation rate (sSFR) for the 94 galaxies published in
  H.-W.\ Chen \etal\ (2010).  We find that the extent of Mg\,II
  absorbing gas, $R_{\rm Mg\,II}$, scales with $M_{\rm star}$ and
  sSFR, following $R_{\rm Mg\,II}\propto M_{\rm star}^{0.28}\times
  {\rm sSFR}^{0.11}$.  The strong dependence of $R_{\rm Mg\,II}$ on
  $M_{\rm star}$ is most naturally explained, if more massive galaxies
  possess more extended halos of cool gas and the observed Mg\,II
  absorbers arise in infalling clouds which will subsequently fuel
  star formation in the galaxies.  The additional scaling relation of
  $R_{\rm Mg\,II}$ with sSFR can be understood either as accounting
  for extra gas supplies due to starburst outflows or as correcting
  for suppressed cool gas content in high-mass halos.  The latter is
  motivated by the well-known sSFR--$M_{\rm star}$ inverse correlation
  in field galaxies.  Our analysis shows that a joint study of
  galaxies and Mg\,II absorbers along common sightlines provides an
  empirical characterization of halo gaseous radius versus halo mass.
  A comparison study of $R_{\rm Mg\,II}$ around red- and blue-sequence
  galaxies may provide the first empirical constraint for resolving
  the physical origin of the observed sSFR--$M_{\rm star}$ relation in
  galaxies.

\end{abstract}


\section{INTRODUCTION}

The Mg\,II $\lambda\lambda\,2796, 2803$ absorption doublets are
commonly seen in photo-ionized gas of temperature $T \sim 10^4$ K
(Bergeron \& Stas\'inska 1986; Charlton et al.\ 2003) and in
high-column density clouds of neutral hydrogen column density $N(\hI)
\approx 10^{18}-10^{22}$ \cmjj\ (Rao et al.\ 2006).  While Mg\,II
absorbers are observed to arise in outflowing gas along the lines of
sight directly into star-forming regions (e.g.\ Weiner \etal\ 2009;
Martin \& Bouch\'e 2009; Rubin \etal\ 2010), it is unclear whether the
absorbers found at projected distances beyond $\rho\sim 10\ h^{-1}$
kpc from star-forming regions are primarily due to outflows, infalling
clouds, or a combination thereof (c.f.\ Bouch\'e \etal\ 2007; M\'enard
\etal\ 2010; Chelouche \& Bowen 2010; Tinker \& Chen 2008; Chen \&
Tinker 2008; Kacprzak \etal\ 2010; Gauthier \etal\ 2010).

The large wavelengths of the doublet transitions, together with
recently available UV sensitive spectrographs on the ground, allow us
to examine in detail the physical connections between the cool gas
probed by Mg\,II absorption features and galaxies routinely found at
redshifts as low as $z\sim 0.1$ (e.g.\ Barton \& Cooke 2009; H.-W.\
Chen \etal\ 2010).  Using a sample of 94 random galaxies at
$z=0.1-0.5$ and $\rho\apll 120\ h^{-1}$ kpc from the line of sight of
a background QSO, H.-W.\ Chen \etal\ (2010; hereafter C10) examined
how the incidence and extent of Mg\,II absorbers depends on galaxy
properties.  They confirmed that the rest-frame absorption equivalent
widths of Mg\,II absorbers, \ewr, decline with increasing projected
distances from the galaxies, and found that the extent of Mg\,II
absorbing gas, $R_{\rm Mg\,II}$, depends sensitively on the galaxy
$B$-band luminosity following $R_{\rm Mg\,II}=75\times
(L_B/L_{B_*})^{(0.35\pm 0.03)}\ h^{-1}$ kpc, but not on the rest-frame
$B_{AB}-R_{AB}$ color or redshift.  The lack of correlation between
\ewr\ and $B_{AB}-R_{AB}$ color suggests a lack of physical connection
between the observed Mg\,II absorbing gas and the recent star
formation history of the galaxies.  It is therefore challenging to
attribute the majority of Mg\,II absorbers found at large galactic
radii ($\apg 10\ h^{-1}$ kpc) to starburst-driven outflows.

However, the strong scaling relation between $R_{\rm Mg\,II}$ and
galaxy $B$-band luminosity, while interesting, is difficult to
interpret.  Although it has been shown that $L_B$ scales with halo
mass $M_h$ (e.g.\ X.\ Yang \etal\ 2005; Tinker \etal\ 2007; Z.\ Zheng
\etal\ 2007), $L_B$ is also found to correlate with [O\,II] luminosity
despite showing a large scatter (G.\ Zhu \etal\ 2009).  It is not
clear whether $L_B$ is a better tracer of halo mass or on-going star
formation.  Here we supplement the galaxy data published in C10 with
new measurements of total stellar mass $M_{\rm star}$ (a more robust
tracer of dark matter halo mass, e.g.\ More \etal\ 2010), \ha\
emission line strength ${\rm EW}(\ha)$, and specific star formation
rate ${\rm sSFR}\equiv {\rm SFR}/M_{\rm star}$ of the galaxies.  These
empirical quantities allow us to investigate possible correlations
between on-going star formation, galaxy mass, and absorber strength,
and to identify whether on-going star formation rate or halo mass is
the dominant factor in determining the incidence and extent of Mg\,II
absorbing gas at large galactic radii.  We adopt a $\Lambda$
cosmology, $\Omega_{\rm M}=0.3$ and $\Omega_\Lambda = 0.7$, with a
dimensionless Hubble constant $h = H_0/(100 \ {\rm km} \ {\rm s}^{-1}\
{\rm Mpc}^{-1})$ throughout the paper.

\begin{figure*}
\begin{center}
\includegraphics[scale=0.45]{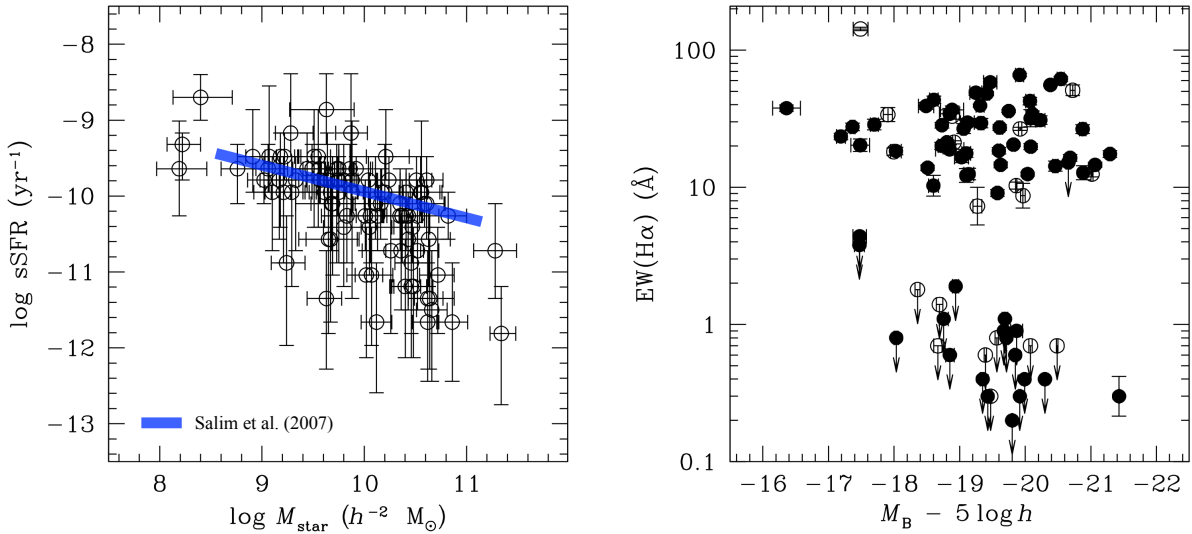}
\caption{General properties of the galaxies in C10.  {\it Left}:
  Comparison of the specific star formation rate (sSFR) and total
  stellar mass estimated from our stellar population synthesis
  analysis.  The errorbars represent the 68\% confidence interval of
  the PDF around the best-fit values.  The plot shows that while the
  uncertainties of the sSFR are large, our galaxies occupy a parameter
  space consistent with what is seen in random field galaxies,
  including the blue-sequence galaxies marked by the solid line
  ($\log\,{\rm sSFR}=-0.35\,(\log\,M_{\rm star}+2\log\,h-9.69)-9.83$)
  from Salim \etal\ (2007) and red galaxies underneath it.  {\it
  Right}: Comparison of \ha\ emission line strength and rest-frame
  $B$-band absolute magnitude, $M_{\rm B}$.  Arrows represent
  2-$\sigma$ upper limits of ${\rm EW}(\ha)$ for galaxies with no \ha\
  detected.  The distribution of data points are also similar to what
  is seen in nearby galaxies (e.g.\ Lee \etal\ 2007). }
\end{center}
\end{figure*}

\section{STELLAR POPULATION SYNTHESIS ANALYSIS}

In order to derive $M_{\rm star}$ and sSFR for each galaxy in our
sample, we follow the maximum likelihood method now routinely adopted
to extract galaxy properties from wide-field survey data (Kauffmann
\etal\ 2003; Brinchmann \etal\ 2004; Salim \etal\ 2005; Walcher \etal\
2008). The basic principle is to calculate the $\chi^2$ between a set
of galaxy observables and equivalent model-galaxy predictions.  The
predictions are derived from a suite of stellar population synthesis
models designed to cover the full range in physical properties of the
observed galaxies (e.g.\ age, star formation history, metallicity,
dust content etc.).

The benefit of this methodology over traditional methods of fitting a
small number of model templates, is in the calculation of a full
probability distribution function (PDF) for each physical property,
which for a set of observables attempts to account for any degeneracy
between derived properties.  In turn, this provides a robust error
estimate for each derived property.  The disadvantages are that the
method relies heavily upon the accuracy of stellar population
synthesis models, and can be somewhat dependent on the prior
distribution of physical properties assumed when creating the library
of model-galaxies.  Stellar population models have advanced greatly in
the last decade, with the development of new empirical and theoretical
stellar libraries and stellar evolutionary tracks (e.g.\ Bruzual
2010).  Results using different model priors have also been compared
and contrasted (e.g.\ Pozzetti \etal\ 2007).  Overall, the concensus
is that $M_{\rm star}$ and sSFR (which in the model is the mean SFR
averaged over the last 10 Myr, divided by $M_{\rm star}$) can be
robustly determined, but may suffer from systematic offsets (by as much
as a factor of two) between the adoption of different population
synthesis models, initial-mass-functions and star formation history
(SFH) priors.

The library of models used in this study is described in detail in da
Cunha \etal\ (2008).  In summary, the stellar population models are
based on those described in Bruzual \& Charlot (2003) revised to
include a new prescription for the TP-AGB evolution of low and
intermediate mass stars (Marigo \& Girardi 2007).
The SFH of the library of model galaxies is parameterized by two
components (Kauffmann \etal\ 2003): an underlying exponentially
declining SFH characterized by an age and decline timescale, and
superimposed random bursts.
The metallicity of the models is distributed
uniformly between 0.02 and 2 times solar. The stellar continuum light
is attenuated according to the two-component dust model of Charlot \&
Fall (2000).
For each model-galaxy in our library we calculate observed-frame
colors at redshifts uniformly distributed between $z=0$ and $z=0.5$,
with a spacing of $\delta z = 0.05$. 

\section{GALAXY PROPERTIES}

All the galaxies in our sample are observed by the SDSS (York \etal\
2000) and have $u$,$g$,$r$,$i$,$z$ photometry.
To obtain accurate colors we use fixed aperture magnitudes with radii
equal to the Petrosian radius\footnote{
http://www.sdss.org/dr7/algorithms/photometry.html} in the r-band.
All magnitudes are corrected for Galactic extinction.
%
%
To identify the best-fit stellar population model for each observed
galaxy, we calculate the $\chi^2$ between the observed optical colors
and derived values of the set of model galaxies at the appropriate
redshift.  We are able to constrain $M_{\rm star}$ and sSFR for all
but three galaxies in the sample.  The results are presented in Table
1.  The distribution of derived $M_{\rm star}$ versus sSFR is shown in
the left panel of Figure 1, along with the best-fit correlation of the
blue-sequence galaxies from Salim \etal\ (2007).  The errorbars
represent the 68\% confidence interval of the PDF around the best-fit
values.  We find that while the uncertainties of the sSFR are large,
our galaxies occupy a parameter space consistent with what is seen in
random field galaxies, including the blue-sequence marked by the solid
line and the red galaxy population underneath it\footnote{We note that
measurements of $M_{\rm star}$ and sSFR remain unchanged within the
errors, when we include in the model analysis available near-IR colors
from the UKIDSS survey (Lawrence et al.\ 2007) and available ${\rm
EW}(\ha)$.  Because these measurements are not available for all
galaxies, we do not include them for consistency.}.


In addition to the broad-band photometric colors, we measure the
rest-frame \ha\ emission equivalent width, ${\rm EW}(\ha)$, using
available galaxy spectra from C10\footnote{We measure ${\rm EW}(\ha)$
instead of \ha\ line fluxes, because the spectra have not been
corrected for aperture or differential slit losses.}.  ${\rm EW}(\ha)$
measures the on-going star formation rate relative to the past average
(e.g.\ Lee \etal\ 2009), and provides an independent, empirical
estimate of the sSFR.  The measurements are presented in Column (5) of
Table 1.  When the \ha\ emission is not observed, we place a
2-$\sigma$ upper limit.  The right panel of Figure 1 shows the
correlation between ${\rm EW}(\ha)$ and the rest-frame $B$-band
absolute magnitude, $M_{\rm B}-5\,\log\,h$, of the galaxies, similar
to what is seen in nearby galaxies (e.g.\ Lee \etal\ 2007).

\begin{figure}
\begin{center}
\includegraphics[scale=0.3]{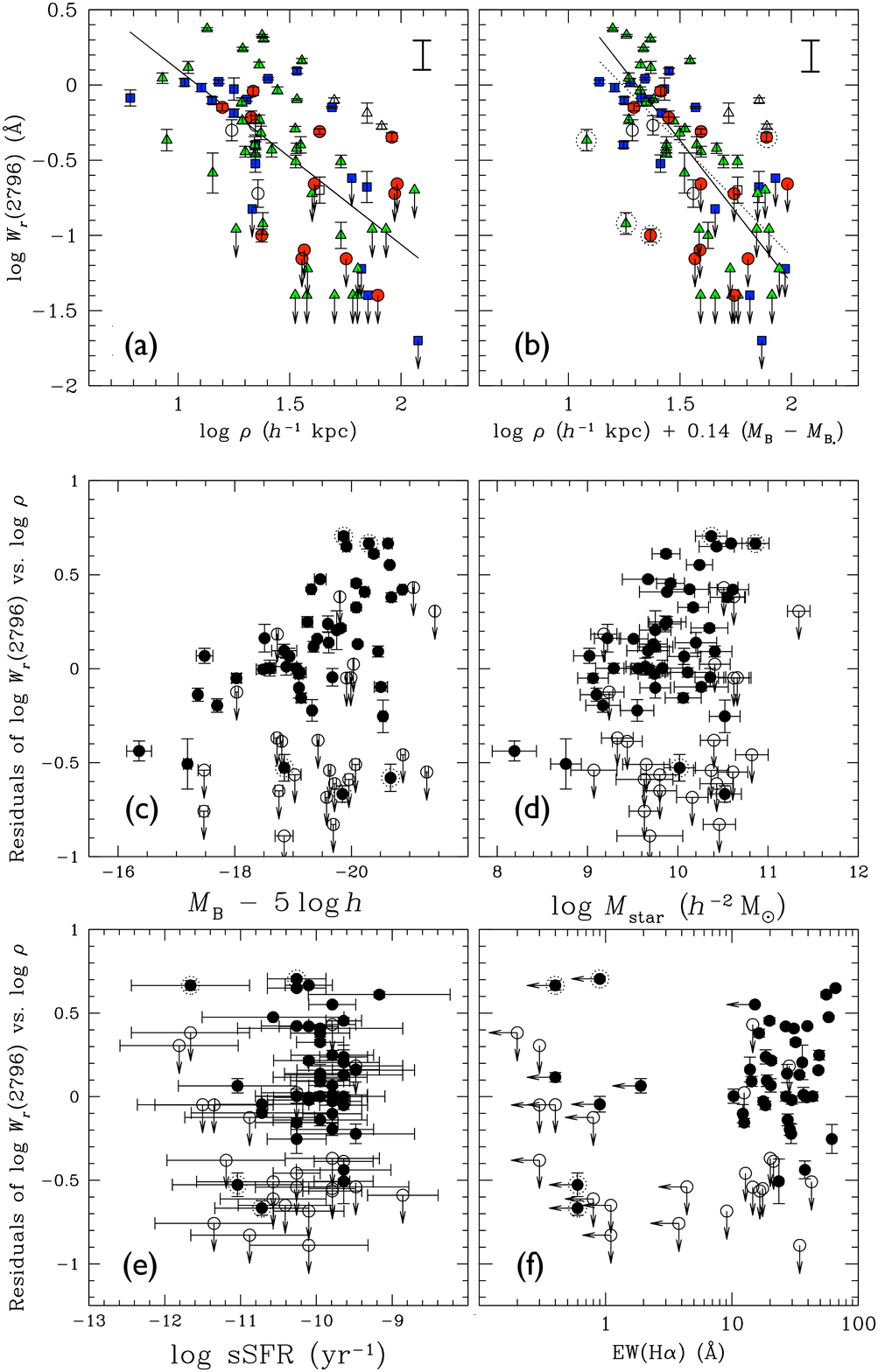}
\caption{The correlation between observed Mg\,II absorption strength,
  \ewr, and different galaxy properties.  The top panels are
  duplicated from C10, showing the \ewr\ versus $\rho$
  anti-correlation before (panel a) and after (panel b) accounting for
  the scaling relation with galaxy $B$-band luminosity.  Circles are
  for galaxies of rest-frame $B_{AB}-R_{AB} > 1.1$; triangles for $0.6
  < B_{AB}-R_{AB} \le 1.1$; and squares for $B_{AB}-R_{AB} < 0.6$.
  Arrows indicate a 2-$\sigma$ upper limit of \ewr, when Mg\,II
  absorption features are not detected.  Open symbols represent
  galaxies with known close neighbors and are not included in the
  model fit.  The solid and dotted lines are the best-fit models,
  excluding and including the outliers (points in dotted circles),
  respectively.  Panels (c)--(f) present the residuals seen in panel
  (a) versus $M_{\rm B}$, $M_{\rm star}$, sSFR, and ${\rm EW}(\ha)$.
  Solid points represent galaxies with detected Mg\,II absorbers,
  while open points represent those galaxies with no detections.  Left
  arrows indicate the 2-$\sigma$ upper limit of ${\rm EW}(\ha)$ when
  the \ha\ emission is not observed.  A comparison between panels (c)
  through (f) suggests that the observed Mg\,II absorber strengths
  depend on galaxy $B$-band magnitude and total stellar mass, but do
  not depend on sSFR or ${\rm EW}(\ha)$.}
\end{center}
\end{figure}

\section{DEPENDENCE OF \ewr\ ON GALAXY PROPERTIES}

To examine the dependence of \ewr\ on galaxy properties, we first
briefly review the findings of C10.  The top panels of Figure 2 are
duplicated from C10.  Panel (a) shows that with a large scatter \ewr\
on average declines with increasing projected distance from the
absorbing galaxy, $\rho$.  Panel (b) of Figure 2 shows that, including
the optimal scaling with galaxy $B$-band luminosity, the \ewr\
vs. $\rho$ anti-correlation is significantly improved.  Different
symbols indicate different rest-frame $B_{AB}-R_{AB}$ color of the
galaxies, showing qualitatively that \ewr\ does not depend sensitively
on galaxy intrinsic colors (see C10 for details).  Roughly 20\% of the
galaxies with $M_{\rm B}-5\,\log\,h\le -18.5$ in this sample have
$B_{AB}-R_{AB}$ color consistent with elliptical or S0.  No clear
trend is seen between $M_B$ and $B_{AB}-R_{AB}$.

\begin{figure*}
\begin{center}
\includegraphics[scale=0.45]{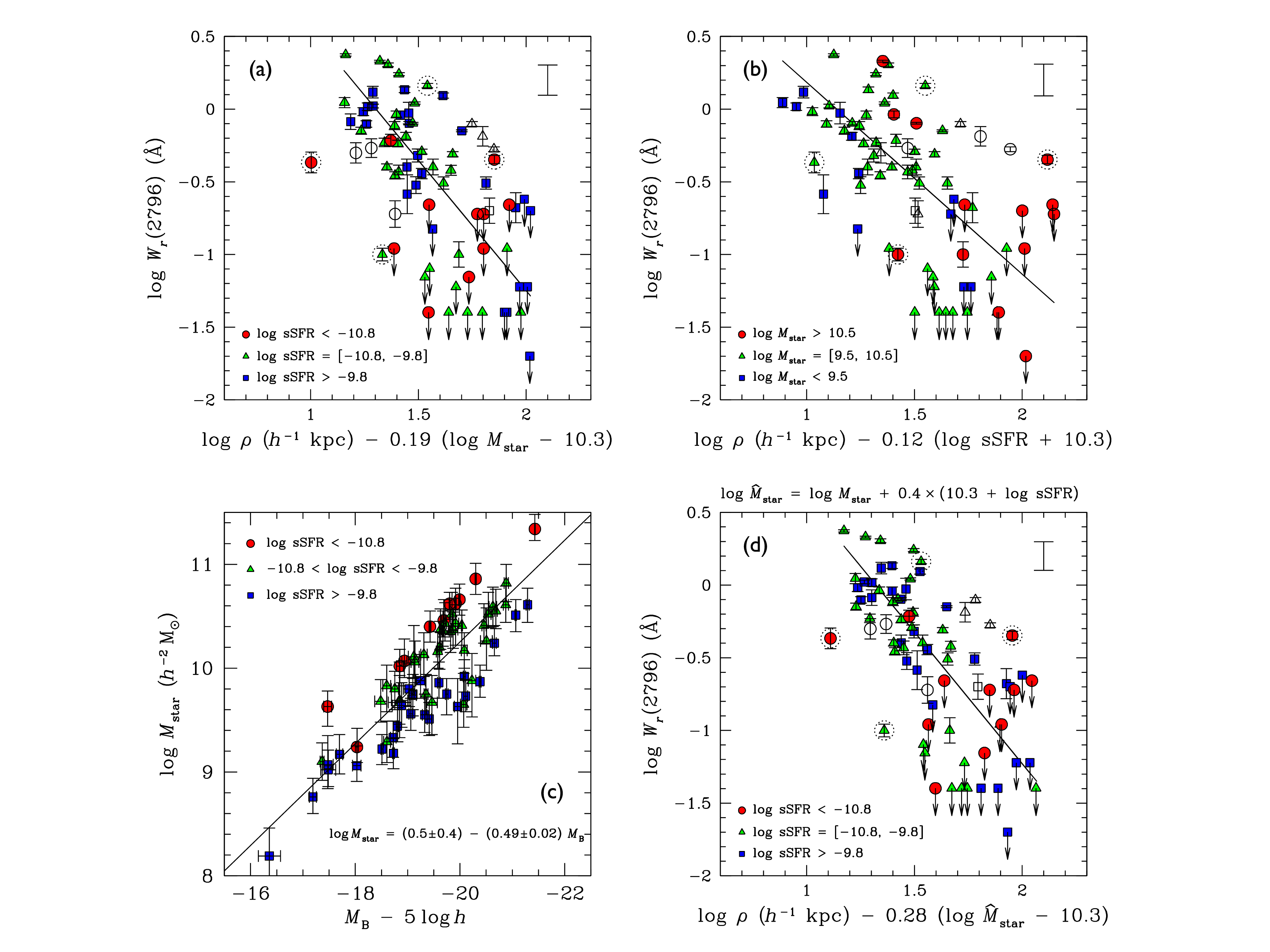}
\caption{Searching for the physical properties that determine the
  incidence and extent of Mg\,II absorbing gas around galaxies.  Panel
  (a) shows the observed \ewr\ vs. $\rho$ anti-correlation, including
  an optimal scaling relation with $M_{\rm star}$.  Different symbols
  represent different ranges in the associated sSFR.  Panel (b) shows
  the observed \ewr\ vs. $\rho$ anti-correlation, including an optimal
  scaling relation with the associated sSFR. Different symbols
  represent different ranges in $M_{\rm star}$.  The same as panels
  (a) and (b) of Figure 1, the open symbols represent galaxies with
  known close neighbors and are excluded from the model fit.  
  Panel (c) shows that while more luminous galaxies have higher
  stellar mass, more massive galaxies also exhibit an on-average lower
  sSFR (see Figure 1 and Salim \etal\ 2007).  Panel (d) shows that
  including sSFR-corrected $M_{\rm star}$ further reduces the scatter
  in the observed \ewr\ vs.\ $\rho$ anti-correlation.}
\end{center}
\end{figure*}

Including new measurements of $M_{\rm star}$, sSFR, and ${\rm
EW}(\ha)$, we present the residuals of the \ewr\ vs.\ $\rho$
anti-correlation (panel a of Figure 2) versus $M_{\rm B}$, $M_{\rm
star}$, sSFR, and ${\rm EW}(\ha)$ in panels (c)--(f) of Figure 2,
respectively.  Inspecting the residuals versus different galaxy
properties, we find that qualitatively the observed Mg\,II absorber
strengths depend sensitively on $M_{\rm B}$ and $M_{\rm star}$, but
not on sSFR or ${\rm EW}(\ha)$.

We also perform the likelihood analysis described in C10, in order to
quantify the best-fit correlation between \ewr\ and different galaxy
properties in the presence of non-detections in the Mg\,II features.
Adopting a power-law model,
\begin{equation}
\log\,\bar{W}_r(2796)\,=\,a_0 + a_1 \log\,\rho + a_2 X,
\end{equation}
we seek the best-fit coefficients, $a_0$, $a_1$, and $a_2$ that
minimize the scatter in the \ewr\ versus $\rho$ anti-correlation.  We
first consider $M_{\rm B}$ and $M_{\rm star}$ separately.  

Assigning $X\equiv\log\,M_{\rm star}-10.3$, we find that the Mg\,II
gaseous extent scales with $\rho$ and $M_{\rm star}$ following
$a_1=-1.8\pm 0.1$ and $a_2=0.34\pm 0.06$ in Equation (1).  For a fixed
\ewr, we find that more massive galaxies possess more extended Mg\,II
absorbing gas.  The extent of Mg\,II absorbing gas at fixed \ewr,
$R_{\rm Mg\,II}$, is found to scale with $M_{\rm star}$ following
$R_{\rm Mg\,II}\propto M_{\rm star}^{0.19\pm 0.03}$.  The results are
presented in panel (a) of Figure 3, where we have also adopted
different symbols to highlight different ranges of sSFR.  The best-fit
model has an associated intrinsic scatter (see C10 for definition) of
$\sigma_c=0.208$ and a r.m.s.\ residual between the observed and model
Mg\,II absorber strengths of ${\rm
r.m.s.}(\log\,W_r-\log\,\bar{W})=0.269$.  We find that, similar to
$M_{\rm B}$, including $M_{\rm star}$ indeed improves the scatter
found in the \ewr\ vs.\ $\rho$ anti-correlation, although the $M_{\rm
star}$-corrected relation exhibits a somewhat larger scatter.  At the
same time, the lack of a systematic trend between the residuals and
sSFR suggests that sSFR has little impact on the extended Mg\,II
absorbing gas at large galactic radii.

Assigning $X\equiv\log\,{\rm sSFR}+10.3$, we find that the Mg\,II
gaseous extent scales with $\rho$ and sSFR following $a_1=-1.3\pm 0.1$
and $a_2=0.15\pm 0.04$ in Equation (1).  The best-fit coefficients
lead to $R_{\rm Mg\,II}\propto {\rm sSFR}^{0.12\pm 0.03}$ at fixed
\ewr, suggesting that galaxies with higher sSFR possess more extended
Mg\,II absorbing gas.  The best-fit model has an associated intrinsic
scatter of $\sigma_c=0.219$ and a r.m.s.\ residual between the
observed and model Mg\,II absorber strengths of ${\rm
r.m.s.}(\log\,W_r-\log\,\bar{W})=0.284$.  Panel (b) of Figure 3 shows
that the remaining scatter in the sSFR-corrected \ewr\ vs.\ $\rho$
anti-correlation is large and there exists an apparent systematic
trend between the residuals of the mean relation and $M_{\rm star}$ as
indicated by different symbols in the panel.

Combining the results shown in panels (a) and (b) of Figure 3, we find
that the scatter in the observed \ewr\ vs.\ $\rho$ anti-correlation is
driven primarily by $M_{\rm star}$ of the galaxies.  However, the
remaining scatter is still somewhat larger than what is seen after
accounting for the scaling relation with galaxy $B$-band luminosity
(panel b of Figure 2), implying that $M_{\rm B}$ is a more successful
than $M_{\rm star}$ in determining the extent of Mg\,II absorbing gas.

To understand the difference, we examine the correlation between
$M_{\rm star}$, $M_{\rm B}$, and sSFR in panel (c) of Figure 3.  We
find that while more luminous galaxies have higher stellar mass, on
average more massive galaxies also exhibit lower sSFR.  Applying a
$\chi^2$ analysis for estimating the best-fit $M_{\rm star}$--sSFR
correlation that reduces the observed scatter, we find that
$\log\,M_{\rm star} \propto (-0.4\pm 0.1) \log\,{\rm sSFR}$ for a
fixed $M_{\rm B}$.  This best-fit correlation is steeper than what is
seen in the blue-sequence galaxies (e.g.\ Salim \etal\ 2007), because
the $\chi^2$ analysis is based on the entire galaxy sample that
includes both red- and blue-sequence galaxies (Figure 1).  

Adopting sSFR-corrected $M_{\rm star}$, $\log\,{\hat M}_{\rm
  star}=\log\,M_{\rm star}+0.4\,(10.3+\log\,{\rm sSFR})$ and
repeating the likelihood analysis to seek the best-fit coefficients in
Equation (1) yield $a_1=-1.8\pm 0.1$ and $a_2=0.5\pm 0.05$, leading to
\begin{equation}
R_{\rm Mg\,II}\propto M_{\rm star}^{0.28\pm 0.03}\times {\rm sSFR}^{0.11\pm 0.03}.
\end{equation}
The results are presented in panel (d) of Figure 3.  The best-fit
model has an associated intrinsic scatter of $\sigma_c=0.195$ and a
r.m.s.\ residual between the observed and model Mg\,II absorber
strengths of ${\rm r.m.s.}(\log\,W_r-\log\,\bar{W})=0.241$.  We find
that including both $M_{\rm star}$ and sSFR reproduces the strong
anti-correlation determined using $M_{\rm B}$ seen in panel (b) of
Figure 2.

\section{DISCUSSION}

Our analysis shows that the extent of Mg\,II absorbing gas depends
strongly on $M_{\rm star}$ and more weakly on sSFR, following $R_{\rm
Mg\,II}\propto M_{\rm star}^{0.28}\times {\rm sSFR}^{0.11}$.  The
strong dependence of \ewr\ on $M_{\rm B}$ (C10) is understood as
$M_{\rm B}$ being an integrated measure of $M_{\rm star}$ and sSFR in
galaxies.  The empirical data demand a scaling relation that includes
both $M_{\rm star}$ and sSFR, with $M_{\rm star}$ being a more
dominant factor.  A tight correlation is found between $M_{\rm star}$
and halo mass (e.g.\ More \etal\ 2010).  The strong dependence between
$R_{\rm Mg\,II}$ and $M_{\rm star}$ is naturally explained, if Mg\,II
absorbers arise in infalling clouds and more massive galaxies possess
more extended halos of cool gas (e.g.\ Mo \& Miralda-Escud\'e 1996;
Maller \& Bullock 2004; Tinker \& Chen 2008).  Our analysis shows that
a joint study of galaxies and Mg\,II absorbers along common sightlines
allows an empirical characterization of halo gaseous radius versus
halo mass.

On the other hand, there are two possible interpretations of the
additional scaling relation between $R_{\rm Mg\,II}$ and sSFR.  The
first scenario is that Mg\,II absorbing gas is produced in wind-blown
materials that, around galaxies of the same mass, can reach to larger
distances at higher SFR.  Likewise, around galaxies of the same SFR
outflows can reach to larger distances in lower-mass halos.  In this
scenario, the gas contents of extended halos around galaxies are
regulated by both accretion and outflows, however, accretion remains
the dominant process.

Alternatively, we consider the inverse correlation between sSFR and
$M_{\rm star}$ commonly seen in field galaxies over a broad redshift
range that has been studied (e.g.\ Figure 1; Salim \etal\ 2007; X.-Z.\
Zheng \etal\ 2007; Y.-M.\ Chen \etal\ 2009).  More massive galaxies on
average exhibit lower sSFR.  In particular, red galaxies show an
overall much reduced rate of star formation in comparison to the
blue-sequence galaxies, suggesting a suppressed gas supply around
these red galaxies (e.g.\ Schiminovich \etal\ 2007).  Instead of
gaseous halos being enriched with starburst driven outflows, the
additional scaling relation between $R_{\rm Mg\,II}$ and sSFR in
Equation (2) can therefore be understood as due to an increasingly
suppressed cool gas supply around higher-mass galaxies.  In this
scenario, Mg\,II absorbers serve as a direct measure of gas supply
around galaxies.

The physical processes that produce the bi-modality of red and blue
galaxies and the tilt of the sSFR--$M_{\rm star}$ relation in the blue
sequence remain uncertain.  While it is generally understood that SFR
follows the net accretion rate of cool gas (e.g.\ Birnboim \& Dekel
2003; Kere{\v s} \etal\ 2005, 2009), achieving galaxy bi-modality in
simulations using feedback, mergers etc.\ does not necessarily
reproduce the time-evolution of the sSFR-$M_{\rm star}$ relation for
the blue sequence galaxies (e.g.\ Dav\'e 2008; Bouch\'e \etal\ 2010).
If observations of Mg\,II absorbers around galaxies give a direct
measure of how the cool halo gas content varies with galaxy mass, a
comparison study of the Mg\,II gaseous extent around red- and
blue-sequence galaxies may provide the first empirical constraint for
resolving the physical origin of the observed sSFR--$M_{\rm star}$
relation in galaxies.

Finally, we note the remaining large scatter in the \ewr\ vs.\ $\rho$
anti-correlation after accounting for the differences in $M_{\rm
star}$ and sSFR.  Given the range of $M_{\rm B}$, $B_{AB}-R_{AB}$,
$M_{\rm star}$, and sSFR covered by the galaxy sample, it is not clear
whether any single physical mechanism can explain such scatter for the
entire sample.  C10 attributed such scatter to Poisson noise in the
number of cool clumps intercepted along a line of sight.  Under a
clumpy halo scenario, we can begin to constrain the size and mass of
individual clumps based on known properties of Mg\,II absorbers.


\acknowledgments

We thank St\'ephane Charlot for providing the updated spectral
library, and Lynn Matthews and Michael Rauch for helpful comments on
an earlier version of the paper.  H.-W.C. acknowledges partial support
from an NSF grant AST-0607510.  VW acknowledges support from a Marie
Curie Intra-European fellowship.  We thank the SDSS collaboration for
producing and maintaining the SDSS public data archive.  Funding for
the SDSS and SDSS-II was provided by the Alfred P. Sloan Foundation,
the Participating Institutions, the National Science Foundation, the
U.S. Department of Energy, the National Aeronautics and Space
Administration, the Japanese Monbukagakusho, the Max Planck Society,
and the Higher Education Funding Council for England. The SDSS was
managed by the Astrophysical Research Consortium for the Participating
Institutions.


\begin{references}

\vskip 0.2in

\reference{} Barton, E. J. \& Cooke, J.\ 2009, AJ, 138, 1817


\reference{} Bergeron, J. \& Stas\'inska, G. 1986, A\&A, 169, 1

\reference{} Birnboim, Y. \& Dekel, A. 2003, MNRAS, 345, 349

\reference{} {Bouch{\'e}}, N., {Murphy}, M.~T., {P{\'e}roux}, C., Davies, R., Eisenhauer, F., F\"orster Schreiber, N. M., \& Tacconi, L. 2007, ApJ, 669, L5

\reference{} Bouch\'e, N., Dekel, A., Genzel, R. \etal\ 2010, ApJ, 718, 1001

\reference{} Brinchmann J., Charlot S., White S.~D.~M., Tremonti C., Kauffmann G., Heckman T., \& Brinkmann J.,  2004, MNRAS, 351, 1151

\reference{} Bruzual, A. G. \& Charlot S. 2003, \mnras, 344, 1000

\reference{} Bruzual, A. G. 2010, in IAU Symposium Vol. 262, p. 55

\reference{} Charlot S. \& Fall S.~M. 2000, \apj, 539, 718

\reference{} Charlton, J. C., Ding, J., Zonak, S. G., Churchill, C. W., Bond, N. A., \& Rigby, J. R. 2003, ApJ, 589, 111

\reference{} {Chelouche}, D. \& {Bowen}, D.~V. 2010, arXiv:1008.2769

\reference{} Chen, H.-W. \& Tinker, J. L.\ 2008, ApJ, 687, 745

\reference{} Chen, H.-W., Helsby, J. E., Gauthier, J.-R., Shectman, S. A., Thompson, I. A., \& Tinker, J. L. 2010, ApJ, 714, 1521 (C10)

\reference{} Chen, Y. M., Wild, V., \& Kauffmann, G. \etal\ 2009, MNRAS, 393, 406


\reference{} da Cunha E., Charlot S., \& Elbaz D. 2008, \mnras, 388, 1595

\reference{} Dav\'e, R. 2008, MNRAS, 385, 147

\reference{} Gauthier, J.-R., Chen, H.-W., \& Tinker, J. L. 2010, ApJ, 716, 1263

\reference{} {Kacprzak}, G.~G., {Churchill}, C.~W., Ceverino, D., {Steidel}, C.~C., Klypin, A., \& {Murphy}, M.~T. 2010, ApJ, 711, 533

\reference{} Kauffmann G.,  Heckman T.~M.,  \& White S.~D.~M., et al.\ 2003, \mnras, 341, 33

\reference{} {Kere{\v s}}, D., {Katz}, N., {Weinberg}, D.~H., \& {Dav{\'e}}, R. 2005, MNRAS, 363, 2

\reference{} {Kere{\v s}}, D., {Katz}, N., Fardel, M., Dav\'e, R., \& {Weinberg}, D. H.\ 2009, MNRAS, 395, 160

\reference{} Lawrence, A. \etal\ 2007, MNRAS, 379, 1599

\reference{} Lee, J. C., Kennicutt, R. C., Funes, S. J., Jos\'e, G., Sakai, S., \& Akiyama, S. 2007, ApJ, 671, L113

\reference{} \ibid\ 2009, ApJ, 692, 1305

\reference{} {Maller}, A.~H. \& {Bullock}, J.~S. 2004, MNRAS, 355, 694

\reference{} Marigo, P. \& Girardi, L. 2007, A\&A, 469, 239

\reference{} Martin, C. L. \& Bouch\'e, N. 2009, ApJ, 703, 1394

\reference{} M\'enard, B., Wild, V., Nestor, D., Quider, A., \& Zibetti, S. 2010, MNRAS submitted (arXiv:0912.3263)

\reference{} {Mo}, H.~J. \& {Miralda-Escud\'e}, J. 1996, ApJ, 469, 589

\reference{} More, S. \etal\ 2010, MNRAS submitted (arXiv:1003.3203)


\reference{} Pozzetti L., Bolzonella M., \& Lamareille F., et al.\ 2007, \aap, 474, 443

\reference{} {Rao}, S.~M., {Turnshek}, D.~A., \& {Nestor}, D.~B. 2006, ApJ, 636, 610

\reference{} Rubin, K. H. \etal\ 2010, ApJ, 719, 1503

\reference{} Salim S. \etal\ 2005, ApJ, 619, L39

\reference{} Salim S. \etal\ 2007, ApJS, 173, 267

\reference{} Schiminovich, D. \etal\ 2007, ApJS, 173, 315

\reference{} Tinker, J. L., Norberg, P., Weinberg, D. H., \& Warren, M. S. 2007, ApJ, 659, 877

\reference{} Tinker, J. L. \& Chen, H.-W. 2008, ApJ, 679, 1218

\reference{} Walcher C.~J., Lamareille F., \& Vergani D., et al.\ 2008, \aap, 491, 713

\reference{} Weiner, B. J. \etal\ 
2009, ApJ, 692, 187

\reference{} Yang, X. \etal\ 2005, MNRAS, 358, 217

\reference{} York, D. G. et al.\ 2000, AJ, 120, 1579

\reference{} Zheng, X. Z., Bell, E., \& Papovich, C. \etal\ 2007, ApJ, 661, L41

\reference{} {Zheng}, Z., {Coil}, A.~L., \& {Zehavi}, I. 2007, ApJ, 667, 760

\reference{} Zhu, G., Moustakas, J., \& Blanton, M. R. 2009, ApJ, 701, 86

\end{references}


\begin{center}
\begin{tiny}
\begin{deluxetable}{p{1.5in}crrr}
\tablewidth{0pc}
\tablecaption{Summary of Additoinal Galaxy Properties in Chen \etal\ (2010)\tablenotemark{a}}
\tabletypesize{\tiny}
\tablehead{\multicolumn{1}{c}{Galaxy ID} & \multicolumn{1}{c}{$z_{\rm spec}$} & 
\multicolumn{1}{c}{$\log\,(M_{\rm star}/h^{2}\,{\rm M_\odot}) $} & 
\colhead{$\log\,{\rm sSFR}/{\rm yr}^{-1}$} & \colhead{${\rm EW(\ha)}$ (\AA)\tablenotemark{b}}\\
\multicolumn{1}{c}{(1)} & \multicolumn{1}{c}{(2)} & \multicolumn{1}{c}{(3)} &
\colhead{(4)} & \colhead{(5)} }
\startdata
SDSSJ003339.85$-$005522.36 & 0.2124 & $ 9.7 \pm 0.2$ & $ -9.6 \pm 0.8$ & $34.2 \pm 1.8$ \nl
SDSSJ003407.78$-$085453.28 & 0.3617 & $ 9.6 \pm 0.2$ & $ -9.8 \pm 0.5$ & $36.8 \pm 2.3$ \nl
SDSSJ003412.85$-$010019.79 & 0.2564 & $10.1 \pm 0.2$ & $-11.0 \pm 0.8$ & $<1.9$ \nl
SDSSJ003414.49$-$005927.49 & 0.1202 & $10.6 \pm 0.1$ & $-11.7 \pm 0.8$ & $<0.2$ \nl
SDSSJ010136.52$-$005016.44 & 0.2615 & $10.4 \pm 0.2$ & $-10.6 \pm 0.7$ & $<0.8$ \nl
SDSSJ010155.80$-$084408.74 & 0.1588 & $ 9.2 \pm 0.2$ & $ -9.8 \pm 0.5$ & $28.7 \pm 2.6$ \nl
SDSSJ010351.82$+$003740.77 & 0.3515 & $ 9.7 \pm 0.2$ & $-10.0 \pm 0.4$ & $18.9 \pm 1.4$ \nl
SDSSJ021558.84$-$011131.23 & 0.2103 & $ 9.8 \pm 0.2$ & $-10.0 \pm 0.6$ & $<0.4$ \nl
SDSSJ022949.97$-$074255.88 & 0.3866 & $ 9.7 \pm 0.3$ & $-10.6 \pm 0.9$ & $58.3 \pm 4.2$ \nl
SDSSJ024127.75$-$004517.04 & 0.1765 & $10.5 \pm 0.2$ & $-11.2 \pm 0.9$ & $<0.6$ \nl
SDSSJ032230.27$+$003712.72 & 0.1833 & $10.2 \pm 0.2$ & $ -9.5 \pm 0.5$ & $12.5 \pm 0.6$ \nl
SDSSJ032232.55$+$003644.68 & 0.2185 & $ 9.0 \pm 0.2$ & $ -9.8 \pm 0.3$ & $20.3 \pm 2.1$ \nl
SDSSJ035241.99$+$001317.13 & 0.3671 & $10.4 \pm 0.2$ & $-10.3 \pm 0.4$ & $<0.9$ \nl
SDSSJ040404.51$-$060709.46 & 0.2387 & $ 9.5 \pm 0.2$ & $ -9.6 \pm 0.3$ & $33.1 \pm 1.7$ \nl
SDSSJ075001.34$+$161301.92 & 0.1466 & $ 8.8 \pm 0.2$ & $ -9.6 \pm 0.4$ & $23.5 \pm 1.3$ \nl
SDSSJ075450.11$+$185005.28 & 0.2856 & $10.5 \pm 0.2$ & $-10.9 \pm 0.8$ & $<1.1$ \nl
SDSSJ075525.13$+$172825.79 & 0.2541 & $10.4 \pm 0.2$ & $-10.1 \pm 0.5$ & $20.5 \pm 0.6$ \nl
SDSSJ080005.11$+$184933.31 & 0.2544 & $ 9.6 \pm 0.2$ & $ -9.5 \pm 0.8$ & $29.4 \pm 2.3$ \nl
SDSSJ082340.56$+$074751.07 & 0.1864 & $10.4 \pm 0.2$ & $-10.7 \pm 0.6$ & $<0.9$ \nl
SDSSJ084120.59$+$012628.85 & 0.4091 & $10.5 \pm 0.2$ & $-10.3 \pm 0.4$ & $61.7 \pm 2.8$ \nl
SDSSJ084455.58$+$004718.15 & 0.1551 & $ 9.8 \pm 0.2$ & $ -9.8 \pm 0.4$ & $12.2 \pm 1.4$ \nl
SDSSJ085829.88$+$022616.04 & 0.1097 & $ 9.4 \pm 0.2$ & $ -9.6 \pm 0.4$ & $21.3 \pm 0.9$ \nl
SDSSJ090519.01$+$084933.70 & 0.3856 & $ 9.7 \pm 0.3$ & $-10.6 \pm 1.0$ & $42.6 \pm 2.7$ \nl
SDSSJ090519.61$+$084932.22 & 0.4545 & $ 9.6 \pm 0.3$ & $ -8.9 \pm 0.5$ & $...$ \nl
SSGAL090519.72$+$084914.02 & 0.1499 & $...$ & $...$ & $37.8 \pm 2.0$ \nl
SDSSJ091845.10$+$060202.93 & 0.1849 & $10.4 \pm 0.2$ & $-11.2 \pm 0.8$ & $<0.3$ \nl
SDSSJ093252.25$+$073731.59 & 0.3876 & $ 9.9 \pm 0.3$ & $-10.0 \pm 1.1$ & $30.9 \pm 2.2$ \nl
SDSSJ093537.25$+$112410.66 & 0.2808 & $ 9.7 \pm 0.2$ & $ -9.8 \pm 0.4$ & $17.7 \pm 1.2$ \nl
SDSSJ100810.61$+$014446.17 & 0.2173 & $10.5 \pm 0.2$ & $ -9.8 \pm 0.3$ & $14.6 \pm 0.6$ \nl
SDSSJ100906.91$+$023557.81 & 0.2523 & $10.5 \pm 0.2$ & $-10.7 \pm 0.6$ & $<0.6$ \nl
SDSSJ102220.71$+$013143.50 & 0.1369 & $10.7 \pm 0.2$ & $-11.5 \pm 0.9$ & $<0.4$ \nl
SDSSJ103605.26$+$015654.88 & 0.3571 & $10.6 \pm 0.2$ & $ -9.8 \pm 0.3$ & $17.5 \pm 1.0$ \nl
SDSSJ103836.38$+$095143.68 & 0.1742 & $ 9.1 \pm 0.2$ & $ -9.6 \pm 0.3$ & $18.4 \pm 1.2$ \nl
SDSSJ112016.63$+$093317.94 & 0.4933 & $10.6 \pm 0.2$ & $-10.1 \pm 0.3$ & $...$ \nl
SDSSJ113756.76$+$085022.38 & 0.3356 & $ 9.9 \pm 0.2$ & $ -9.8 \pm 0.5$ & $49.0 \pm 2.1$ \nl
SDSSJ114144.83$+$080554.09 & 0.2290 & $ 9.9 \pm 0.2$ & $ -9.6 \pm 0.3$ & $18.5 \pm 1.0$ \nl
SDSSJ114145.14$+$080605.27 & 0.3583 & $10.2 \pm 0.2$ & $-10.0 \pm 0.3$ & $31.9 \pm 4.6$ \nl
SDSSJ120931.61$+$004546.23 & 0.2533 & $ 9.8 \pm 0.2$ & $-10.4 \pm 0.6$ & $<1.1$ \nl
SDSSJ122115.84$-$020259.37 & 0.0934 & $ 8.2 \pm 0.2$ & $ -9.3 \pm 0.3$ & $...$ \nl
SDSSJ125737.93$+$144802.20 & 0.4648 & $...$ & $...$ & $...$ \nl
SDSSJ130555.49$+$014928.62 & 0.2258 & $10.2 \pm 0.2$ & $-10.1 \pm 0.5$ & $9.1 \pm 0.5$ \nl
SDSSJ130557.05$+$014922.34 & 0.1747 & $10.9 \pm 0.2$ & $-11.7 \pm 0.8$ & $<0.4$ \nl
SDSSJ132757.22$+$101136.02 & 0.2557 & $ 9.3 \pm 0.2$ & $-10.0 \pm 0.9$ & $43.5 \pm 2.7$ \nl
SDSSJ132831.54$+$075943.00 & 0.3323 & $10.4 \pm 0.2$ & $-10.0 \pm 0.3$ & $14.3 \pm 1.2$ \nl
SDSSJ132832.74$+$075952.56 & 0.2358 & $ 9.8 \pm 0.2$ & $ -9.6 \pm 0.5$ & $36.0 \pm 0.8$ \nl
SDSSJ133905.86$+$002225.36 & 0.1438 & $10.6 \pm 0.2$ & $-11.4 \pm 0.8$ & $<0.3$ \nl
SDSSJ140618.34$+$130143.61 & 0.1748 & $10.4 \pm 0.2$ & $-10.3 \pm 0.5$ & $12.5 \pm 0.4$ \nl
SDSSJ140619.94$+$130105.23 & 0.2220 & $ 9.6 \pm 0.2$ & $ -9.6 \pm 0.3$ & $26.8 \pm 2.2$ \nl
SDSSJ142600.05$-$001818.12 & 0.1382 & $11.3 \pm 0.1$ & $-11.8 \pm 0.8$ & $0.3 \pm 0.1$ \nl
SDSSJ143216.97$+$095522.23 & 0.3293 & $10.1 \pm 0.2$ & $-10.3 \pm 0.5$ & $39.5 \pm 1.0$ \nl
SDSSJ150339.62$+$064235.04 & 0.2333 & $ 9.3 \pm 0.2$ & $ -9.8 \pm 0.6$ & $20.1 \pm 1.2$ \nl
SDSSJ150340.15$+$064308.11 & 0.1809 & $ 9.6 \pm 0.2$ & $-11.4 \pm 0.8$ & $<3.8$ \nl
SDSSJ151228.25$-$011216.09 & 0.1284 & $ 9.2 \pm 0.1$ & $ -9.5 \pm 0.3$ & $13.9 \pm 0.6$ \nl
SDSSJ153112.77$+$091119.72 & 0.3265 & $ 9.8 \pm 0.2$ & $ -9.8 \pm 0.4$ & $16.6 \pm 1.5$ \nl
SDSSJ153113.01$+$091127.02 & 0.2659 & $ 9.8 \pm 0.2$ & $-10.3 \pm 0.5$ & $10.3 \pm 1.8$ \nl
SDSSJ153715.67$+$023056.39 & 0.2151 & $ 9.5 \pm 0.2$ & $ -9.5 \pm 0.8$ & $48.3 \pm 1.6$ \nl
SDSSJ155336.77$+$053438.23 & 0.3227 & $10.2 \pm 0.1$ & $ -9.8 \pm 0.3$ & $<15.2$ \nl
SDSSJ155556.54$-$003615.58 & 0.3006 & $ 9.7 \pm 0.4$ & $-10.1 \pm 0.8$ & $34.4 \pm 3.0$ \nl
SDSSJ160749.54$-$002228.42 & 0.3985 & $10.6 \pm 0.2$ & $-10.1 \pm 0.4$ & $26.6 \pm 2.0$ \nl
SDSSJ160906.36$+$071330.66 & 0.2075 & $10.4 \pm 0.2$ & $-10.3 \pm 0.5$ & $14.6 \pm 0.7$ \nl
SDSSJ204303.53$-$010139.05 & 0.2356 & $ 9.9 \pm 0.2$ & $ -9.2 \pm 0.9$ & $55.8 \pm 0.7$ \nl
SDSSJ204304.34$-$010137.91 & 0.1329 & $ 9.2 \pm 0.2$ & $-10.9 \pm 0.9$ & $<0.8$ \nl
SDSSJ210230.86$+$094121.06 & 0.3565 & $10.1 \pm 0.3$ & $-10.1 \pm 0.5$ & $29.7 \pm 2.1$ \nl
SDSSJ212938.98$-$063758.80 & 0.2782 & $ 9.7 \pm 0.2$ & $-10.1 \pm 0.5$ & $39.2 \pm 2.2$ \nl
SDSSJ221126.42$+$124459.93 & 0.4872 & $10.3 \pm 0.3$ & $-10.7 \pm 0.9$ & $...$ \nl
SDSSJ221526.04$+$011353.78 & 0.3203 & $10.2 \pm 0.2$ & $-10.0 \pm 0.5$ & $27.3 \pm 1.3$ \nl
SDSSJ221526.88$+$011347.20 & 0.1952 & $ 9.1 \pm 0.3$ & $ -9.5 \pm 0.7$ & $<4.4$ \nl
SDSSJ222849.01$-$005640.04 & 0.2410 & $10.0 \pm 0.2$ & $-10.3 \pm 0.5$ & $ 7.3 \pm 2.3$ \nl
SDSSJ223246.44$+$134655.34 & 0.3221 & $10.6 \pm 0.2$ & $-10.0 \pm 0.4$ & $16.4 \pm 0.9$ \nl
SDSSJ223316.34$+$133315.37 & 0.2138 & $ 9.9 \pm 0.2$ & $ -9.6 \pm 0.2$ & $19.8 \pm 1.0$ \nl
SDSSJ223359.74$-$003320.83 & 0.1162 & $ 9.1 \pm 0.2$ & $-10.0 \pm 0.6$ & $27.6 \pm 1.7$ \nl
SDSSJ224704.01$-$081601.00 & 0.4270 & $10.8 \pm 0.2$ & $-10.3 \pm 0.3$ & $12.8 \pm 1.5$ \nl
SDSSJ230225.06$-$082156.65 & 0.3618 & $10.4 \pm 0.2$ & $-10.3 \pm 0.4$ & $66.0 \pm 1.5$ \nl
SDSSJ230225.17$-$082159.07 & 0.2146 & $ 8.4 \pm 0.3$ & $ -8.7 \pm 0.3$ & $142.9 \pm 4.3$ \nl
SDSSJ230845.53$-$091445.97 & 0.2147 & $10.0 \pm 0.2$ & $-11.0 \pm 0.9$ & $<0.6$ \nl
SDSSJ232812.79$-$090603.73 & 0.1148 & $ 9.2 \pm 0.1$ & $ -9.5 \pm 0.6$ & $28.4 \pm 1.3$ \nl
SDSSJ234949.42$+$003542.34 & 0.2778 & $10.1 \pm 0.2$ & $-10.3 \pm 0.5$ & $12.5 \pm 1.2$ \nl
\hline
\multicolumn{5}{c}{"Group" Galaxies} \\
\hline
SDSSJ003339.66$-$005518.36 & 0.1760 & $10.1 \pm 0.2$ & $-11.7 \pm 0.9$ & $<1.8$ \nl
SDSSJ003341.47$-$005522.79 & 0.1758 & $ 8.9 \pm 0.2$ & $ -9.5 \pm 0.5$ & $33.9 \pm 4.0$ \nl
SDSSJ074527.22$+$192003.88 & 0.4582 & $10.6 \pm 0.2$ & $-10.0 \pm 0.6$ & $51.0 \pm 4.6$ \nl
SDSSJ083218.55$+$043337.81 & 0.1681 & $10.4 \pm 0.2$ & $-10.3 \pm 0.5$ & $10.3 \pm 0.6$ \nl
SDSSJ083218.77$+$043346.58 & 0.1678 & $10.7 \pm 0.2$ & $-11.0 \pm 0.7$ & $<0.7$ \nl
SDSSJ083221.60$+$043359.74 & 0.1693 & $ 9.3 \pm 0.2$ & $ -9.2 \pm 0.5$ & $21.3 \pm 0.6$ \nl
SDSSJ091845.70$+$060220.57 & 0.7967 & $...$ & $...$ & $...$ \nl
SDSSJ100807.63$+$014443.39 & 0.3290 & $ 9.6 \pm 0.2$ & $ -9.8 \pm 0.3$ & $...$ \nl
SDSSJ114830.94$+$021807.91 & 0.3206 & $10.6 \pm 0.2$ & $-10.6 \pm 0.5$ & $8.7 \pm 1.8$ \nl
SDSSJ121347.09$+$000141.26 & 0.2258 & $ 9.6 \pm 0.2$ & $-10.0 \pm 0.5$ & $<0.7$ \nl
SDSSJ121347.14$+$000136.62 & 0.2259 & $10.6 \pm 0.2$ & $-11.4 \pm 0.9$ & $<0.8$ \nl
SDSSJ132829.30$+$080003.17 & 0.2549 & $10.0 \pm 0.2$ & $ -9.8 \pm 0.3$ & $26.7 \pm 0.7$ \nl
SDSSJ132830.62$+$080005.22 & 0.2537 & $10.5 \pm 0.2$ & $-10.4 \pm 0.5$ & $<0.7$ \nl
SDSSJ132831.15$+$075923.90 & 0.2537 & $10.1 \pm 0.2$ & $-10.4 \pm 0.5$ & $<1.4$ \nl
SDSSJ153717.42$+$023026.46 & 0.3114 & $11.3 \pm 0.2$ & $-10.7 \pm 0.6$ & $...$ \nl
SDSSJ204431.32$+$011304.97 & 0.1927 & $10.5 \pm 0.2$ & $-11.2 \pm 0.8$ & $<0.3$ \nl
SDSSJ204431.87$+$011308.81 & 0.1921 & $ 9.2 \pm 0.2$ & $-10.0 \pm 0.7$ & $18.2 \pm 1.1$ \nl
\enddata
\tablenotetext{a}{SDSSJ125737.93$+$144802.20 is a broad-line AGN;
  SSGAL090519.72$+$084914.02 is heavily blended with the background
  QSO in the SDSS images; and SDSSJ091845.70$+$060220.57 is found in
  the QSO host environment.  We therefore do not have constraints for
  their stellar mass or sSFR.}  \tablenotetext{b}{No detections are
  expressed as 2-$\sigma$ upper limits to the underlying emission line
  strength of warm ISM.  Galaxies with no spectral coverage of the
  \ha\ emission are indicated with '...'.}
\end{deluxetable}
\end{tiny}
\end{center}

\end{document}